# Towards protein folding pathways by reconstructing protein residue networks with a policy-driven model

Susan Khor[†]

6 April, 2026

A method that reconstructs protein residue networks using suitable node selection and edge recovery policies produced numerical observations that correlate strongly (Pearson's correlation coefficient < -0.83) with published folding rates for 52 two-state folders and 21 multi-state folders; correlations are also strong at the fold-family level. These results were obtained serendipitously with the ND model, which was introduced previously, but is here extended with policies that dictate actions according to feature states. This result points to the importance of both the starting search point and the prevailing condition (random seed) for the quick success of policy search by a simple hill-climber. The two conditions, suitable policies and random seed, which (evidenced by the strong correlation statistic) setup a conducive environment for modelling protein folding within ND, could be compared to appropriate physiological conditions required by proteins to fold naturally. Of interest is an examination of the sequence of restored edges for potential as plausible protein folding pathways. Towards this end, trajectory data is collected for analysis and further model evaluation and development.

**Background**

A Protein Residue Network (PRN) is a set of nodes positioned at the CA coordinates of a native state protein chain, plus a set of undirected edges connecting node pairs that satisfy certain conditions. A Short Cut Network (SCN) is a unique sub-network of a PRN, made up of edges identified by the Euclidean Distance Search (EDS) algorithm. EDS is a greedy local search algorithm that exploits the small-world character of PRNs to find paths between ordered node pairs. The edges of a SCN are called shortcuts because their presence obviate the need for EDS to do backtracking.

The Network Dynamics (ND) method models protein folding by *re*constructing PRNs. The ND method begins with an edgeless PRN whose nodes are fixed at their native-state positions. It then proceeds to restore the PRN's edges by iterating the following two steps. First, ND chooses a node according to a node selection policy (*ndv*). Second, ND attempts to restore a chosen node's edges by an edge recovery policy (*nde*).

As a PRN is *re*constructed by the ND method, the EDS algorithm is run on variously partially completed PRNs and identifies different sets of shortcuts edges. Shortcuts identified on partially completed PRNs may not all belong to the SCN of the completed PRN. Shortcut edges identified by EDS forms the basis of the "energy" calculation or valuation of a SCN. Native shortcuts, those that appear in the SCN of the completed PRN, decrease the "energy" of a SCN. Conversely, non-native shorts, those that do not appear in the SCN of the completed PRN, increase the valuation of a SCN. A SCN exhibits peak energy (peak*E*) when its observed valuation is highest.

The challenge is to coax the ND method, by finding suitable *ndv* and *nde* policies, to reconstruct a set of PRNs such that their SCNs reach peak energies that correlate strongly with published folding rates. This correlation is a negative one: larger folding rates (faster folders) associate with smaller peak*E* values, and vice versa. For two-state folders, higher peak SCN energies point to larger barriers that slow down the folding process modelled as PRN *re*construction.

---

[†] E-mail: slc.khor@gmail.com

PRN construction, SCN identification by the EDS algorithm, and the ND method were introduced previously in reference [1]. That study focused on two-state folders, used a simple edge recovery strategy, and reported poor results for multi-state folders. This study attends to fold-family level results and reports on how a contextualized edge restoration probability can significantly improved ND results for both two- and multi-state folders.

**Dataset**

Protein chains are partitioned into three fold-families by their simplified DSSP [2] assigned secondary structure; each chain residue is labelled **H** (DSSP: G, H, I), **S** (DSSP: E, B), or **T** (otherwise). The three fold-families are: **A** for chains with predominantly α-helix secondary structure (number of α residues (nH) > 3, number of β residues (nS) < 11, and nH/nS ≈ nH or >> 1); **B** for chains with predominantly β-strand secondary structure (nH < 4, nS > 10, and nH/nS ≈ 0.0); and **AB** otherwise.

Using the above criteria, the Uzunoglu dataset [3] consists of 15 **A** (1ayi, 1enh, 1fex, 1idy, 1imq, 1lmb, 1ryk, 1ss1, 1w4e, 1w4j, 1yza, 2abd, 2bth, 2pdd, 1hrc), 17 **B** (1c9o, 1csp, 1e0l, 1e0m, 1fnf_9, 1g6p, 1jo8, 1k9q, 1m9s, 1mjc, 1psf, 1rlq, 1shf, 1shg, 1ten, 1tit, 3ait), and 20 **AB** (1afi, 1aps, 1bf4, 1div_c, 1div_n, 1fkb, 1n88, 1o6x, 1pgb, 1poh, 1rfa, 1ris, 1spr, 1ubq, 1urn, 2acy, 2ci2, 2ptl, 2qjl, 2vik) two-state folders with lengths (number of amino acids) ranging from 37 to 126. And the Kamagata dataset [4] comprises 5 **A** (2cro, 1cei, 2abd, 1dwr, 2a5e), 3 **B** (1hng, 1tit, 1ttg), and 13 **AB** (1pgb, 1gxt, 1bta, 1bni, 1adw, 3chy, 1ael, 1joo, 1ilb, 2rn2, 1ra9, 1l63, 1php) multi-state folders with lengths ranging from 56 to 175.

**Policy-driven ND**

The ND method is parameterized by two policies, *ndv* and *nde*, which govern how nodes are selected and how edges are recovered, respectively. The basic idea is to select an *incomplete node*, i.e. a node with missing or hitherto unrecovered edges, and then attempt to complete the node by restoring all its PRN edges.

The node selection policy (*ndv*) is determined by fold-family. For both A and B fold-families, *ndv* chooses uniformly at random from incomplete nodes. For the AB fold-family, *ndv* does *crony selection*: if this is the first node selection, choose uniformly at random from all nodes; otherwise choose uniformly at random from incomplete nodes that are directly connected to the previously chosen node. If all nodes within the direct neighbourhood of the previously chosen node (includes the previously chosen node itself) are complete, then choose uniformly at random from all remaining incomplete nodes. In contrast to uniform selection, crony selection attempts to localize edge recovery. From previous experience [1], this approach seems favourable for AB chains.

The edge recovery policy (*nde*) prescribes an *action* given an *edge-state*. To follow the protein folding principle of minimizing loss in chain entropy, the edge recovery process is biased towards restoring edges whose range is more local on the linear protein sequence. Edges with smaller sequence separation tend to have higher edge restoration probability ($p_e$) than edges with larger sequence separation. The four possible actions prescribed by *nde*, **s** (scaled), **sl** (scaled-lower), **su** (scaled-upper), and **u** (unscaled), tunes the extent of this bias (Table 1). Actions **s** and **u** exert the strongest and the weakest bias, respectively (Fig. 1).

**Table 1** Definitions of the four *nde* actions

| Action | Description | Edge restoration probability ($p_e$) for an edge ($u$, $v$) |
|---|---|---|
| **u** | unscaled | $p_e^{unscaled} = 1.0 / \ln(|u-v|)$ |
| **s** | scaled | $p_e^{scaled}$ = maximum of 1e-4 and ($p_e^{unscaled} - p_e^{min}$) / ($p_e^{max} - p_e^{min}$) |
| **sl** | scaled-lower | $p_e^{scaled-lower} = p_e^{scaled}$ plus a small positive adjustment δ where δ is a uniform random real between 1e-3 and ($p_e^{unscaled} - p_e^{scaled}$) /2.0 |
| **su** | scaled-upper | $p_e^{scaled-upper} = p_e^{scaled}$ plus a small positive adjustment δ where δ is a uniform random real between ($p_e^{unscaled} - p_e^{scaled}$) /2.0 and ($p_e^{unscaled} - p_e^{scaled}$) |

In Table 1, |$u$-$v$| is the sequence separation between nodes $u$ and $v$; $p_e^{min}$ and $p_e^{max}$ are the inverse of the natural logarithm (ln) of the largest and smallest sequence separation for a PRN, respectively. The bias effect of each *nde* action on $p_e$ are compared in Fig. 1 for a range of hypothetical sequence distances.

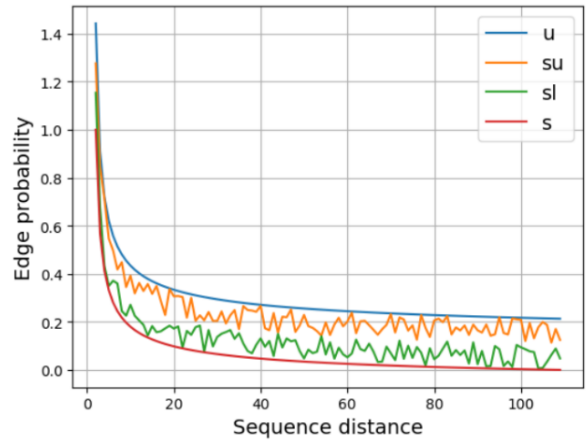

**Fig. 1** Edge restoration probability ($p_e$) under the influence of *nde* actions.

The concept of edge-state captures an edge's secondary structure context within the direct (1-hop) neighbourhood of the edge's two end-point nodes. PRN nodes are initially tagged as **H** (DSSP: G, H, I), **S** (DSSP: E, B), or **T** (otherwise). *N1SS$_v$* labelling expands on these initial tags to include the tags within a node's 1-hop neighbourhood. For example, the *N1SS$_v$* label for an **S** tagged node linked directly to six nodes tagged **H** and two nodes tagged **T**, is HST. The set of all possible *N1SS$_v$* labels are H, S, T, HS, HT, ST, and HST.

The edge-state for edge ($u$, $v$) is the union of the *N1SS$_v$* labels of nodes $u$ and $v$. In Fig. 2, since the green node and the yellow node have the *N1SS$_v$* labels of HS and HST respectively, the edge-state of the link between these two nodes is HS-HST.

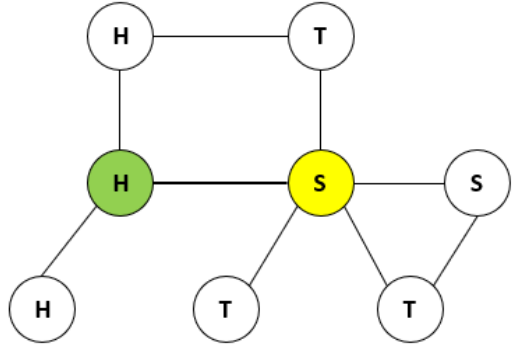

**Fig. 2** Edge-state of the bolded link is HS-HST.

Observe that the *N1SS$_v$* labels of an edge's end-point nodes must have at least one tag (**H**, **S** or **T**) in common. For instance, H-S is not a valid edge-state because there can be no edge that connects a node whose *N1SS$_v$* label is H (which means the **H** tagged node is directedly connected to **H** tagged nodes only) to a node whose *N1SS$_v$* label is S (which means the **S** tagged node is directly connected to **S** tagged nodes only). Hence, the set of possible edge-states are: H-H, H-HS, H-HT, H-HST, S-S, S-HS, S-ST, S-HST, T-T, T-HT, T-ST, T-HST, HS-HS, HS-HT, HS-ST, HS-HST, HT-HT, HT-ST, HT-HST, ST-ST, ST-HST, and HST-HST.

**Policy Search**

Given a set of PRNs and the node selection policy (*ndv*) described above, the objective is to find an edge recovery policy, denoted *nde\**, that optimizes the correlation between peak*E* values produced by ND and published folding rates.

With 22 possible edge-states and four possible actions per edge-state, the size of the search space is $4^{22}$ or $10^{13.24532}$. However, not all 22 edge-states appear in every set of PRNs. For example, the five A fold-family PRNs and the three B fold-family PRNs from the Kamagata dataset, have only eight and five relevant edge-states, respectively. Edge-states that do not appear in the set of given PRNs are discovered prior to search, ignored during search, and prescribed no action by *nde*.

The search for *nde\** is performed by a simple greedy hill-climber, starting at a best single-action *nde* policy. A single-action *nde* policy prescribes the same action for all relevant edge-states. The four possible single-action *nde* policies are evaluated, and the one which produced the strongest correlation becomes "the parent genotype" or starting point of search for the hill-climber.

In hill-climbing search, the "genes" of the parent genotype are mutated (actions prescribed for relevant edge-states are altered) to create "child genotypes" (possibly new *nde* policies). If a child genotype is found to be fitter (is a better *nde* policy) than its parent, it replaces the parent genotype, which is the current *nde\**, and the hill-climbing search continues with the new parent genotype.

The hill-climber used to find *nde\** in this report does an exhaustive and systematic scan of the 1-mutation neighbourhood before proceeding to the 2-mutation neighbourhood. The search is stopped after 204 distinct *nde* policies have been evaluated.

Fig. 3 illustrates a hill-climbing search by fold-family for *nde\** with the Kamagata dataset. The exploration phase determined that the initial parent genotype for **A**, **B** and **AB** fold-families are single-action policies with actions **sl**, **u**, and **u**, respectively (Fig. 3 top). In the exploitation phase, for each fold-family, the hill-climber generates new policy variants from the current parent genotype, evaluates them, and replaces the current parent genotype if a better policy variant is found. The red x's in Fig. 3 (middle) mark when the best *nde* policies (*nde*\*) are found for fold-families **A**, **B** and **AB**, respectively, within the allotted search resources, i.e. number of unique *nde* policy evaluations. The peak$E$ values produced by ND with *nde*\* for each fold-family are plotted against published folding rates in Fig. 3 (bottom). The fold-family specific Pearson's correlation coefficients are all below -0.85.

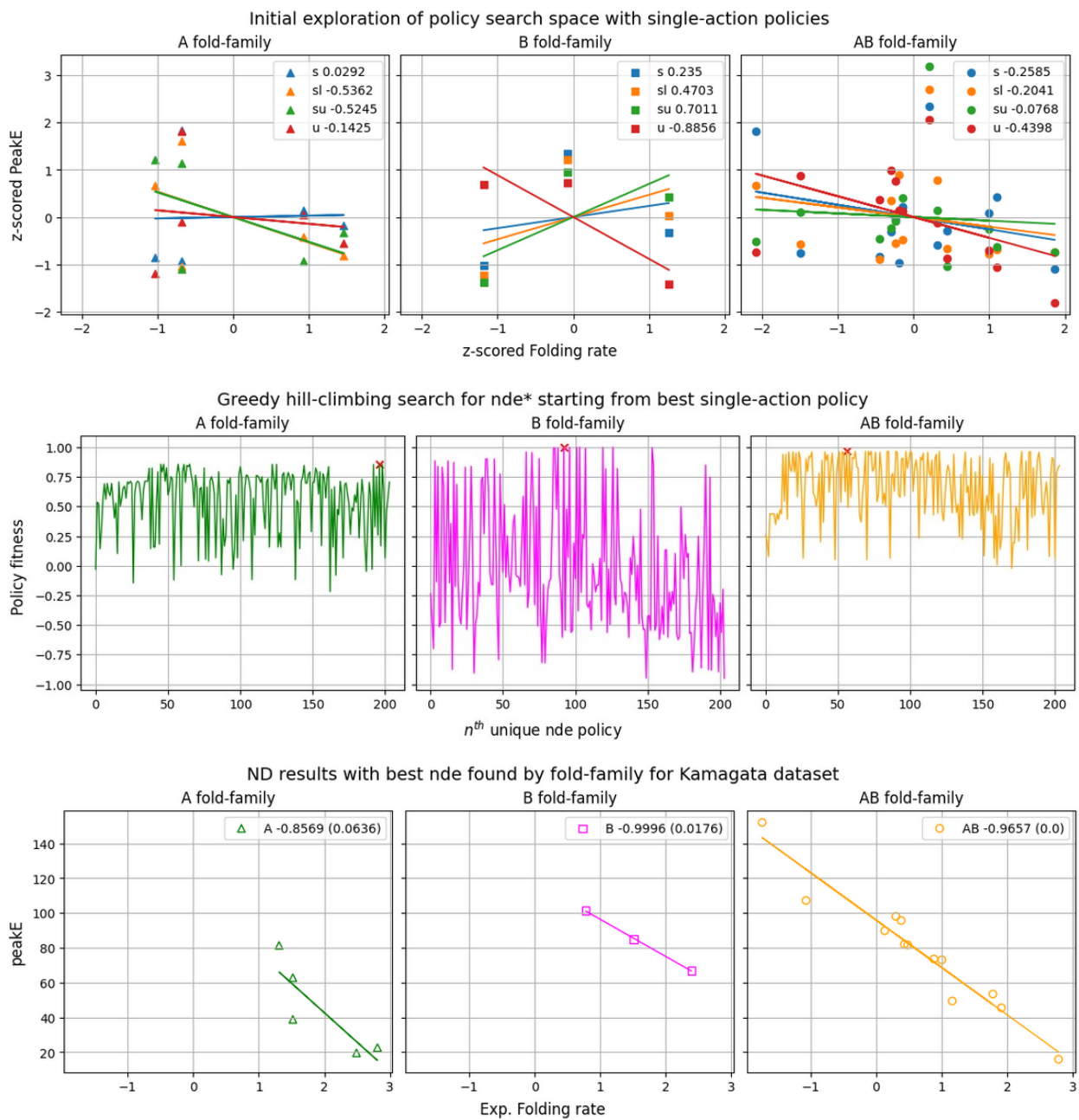

**Fig. 3** A hill-climbing search for *nde*\* by fold-family with the Kamagata dataset.

**Results and Discussion**

The peak*E* values generated by ND with *nde** for each fold-family are put together to produce the overall correlation with published folding rates for the two datasets, as shown in Figs. 4 and 5. Coincidentally, this simple combination yielded strong (ρ < -0.83) and statistically significant (p-value ≈ 0.0) Pearson's correlation coefficients for both datasets. The contributing *nde* policies are inscribed in Tables 2 and 3.

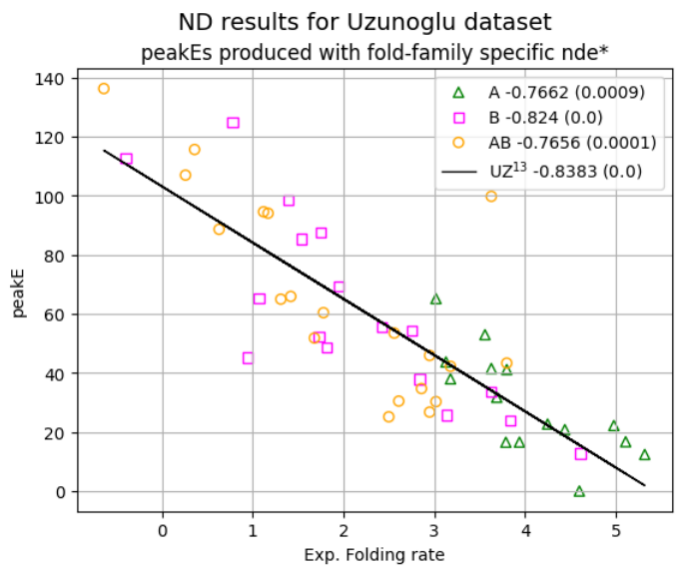

**Fig. 4** ND results with random seed 13 and *nde** found by fold-family for the 52 two-state folders in the Uzunoglu dataset. The combined Pearson's correlation coefficient is -0.8383 with p-value ≈ 0.0. Fold-family specific correlations are all below -0.765 and significant with p-value < 0.001.

**Table 2.** The best *nde* policies found by fold-family A, B and AB supporting the ND results in Fig. 4.

| Edge-state | H-H | H-HS | H-HT | H-HST | S-S | S-HS | S-ST | S-HST | T-T | T-HT | T-ST | T-HST | HS-HS | HS-HT | HS-ST | HS-HST | HT-HT | HT-ST | HT-HST | ST-ST | ST-HST | HST-HST |
|---|---|---|---|---|---|---|---|---|---|---|---|---|---|---|---|---|---|---|---|---|---|---|
| A | u | - | s | - | - | - | - | - | sl | sl | s | su | - | - | - | - | sl | su | sl | sl | su | su |
| B | - | - | - | - | sl | u | sl | u | su | - | su | sl | - | - | su | s | - | - | - | sl | sl | su |
| AB | u | sl | u | sl | su | su | su | su | s | su | sl | su | sl | sl | su | sl | su | su | u | sl | sl | s |

The two-step procedure described above, first obtain good peak*E* values by fold-family and then combine them, is essential to produce the strong overall correlation. *Ceteris paribus* and with the same random seed used to produce the result in Fig. 5, the hill-climbing method searched for a universal *nde**, i.e. one *nde** for all chains in the Kamagata dataset. The outcome of this experiment is displayed on the right in Fig. 5: the overall correlation is much weaker (ρ < -0.66), and so are the correlations at the fold-family level. This poorer result could be due to conflicting best policies between fold-families. Additionally, almost 62% (13/21) of chains in this dataset belong to the AB fold-family. This dataset imbalance would bias the policy search, during both the exploration and exploitation phases, to optimize for the dominating fold-family. Further, searching for a universal *nde** enlarges the search space for some sets of PRNs unnecessarily. ND is native-centric; PRNs are built from coordinates of native chain topologies [5]. Therefore, secondary structure is known and fold-family can be computed with only a little extra effort.

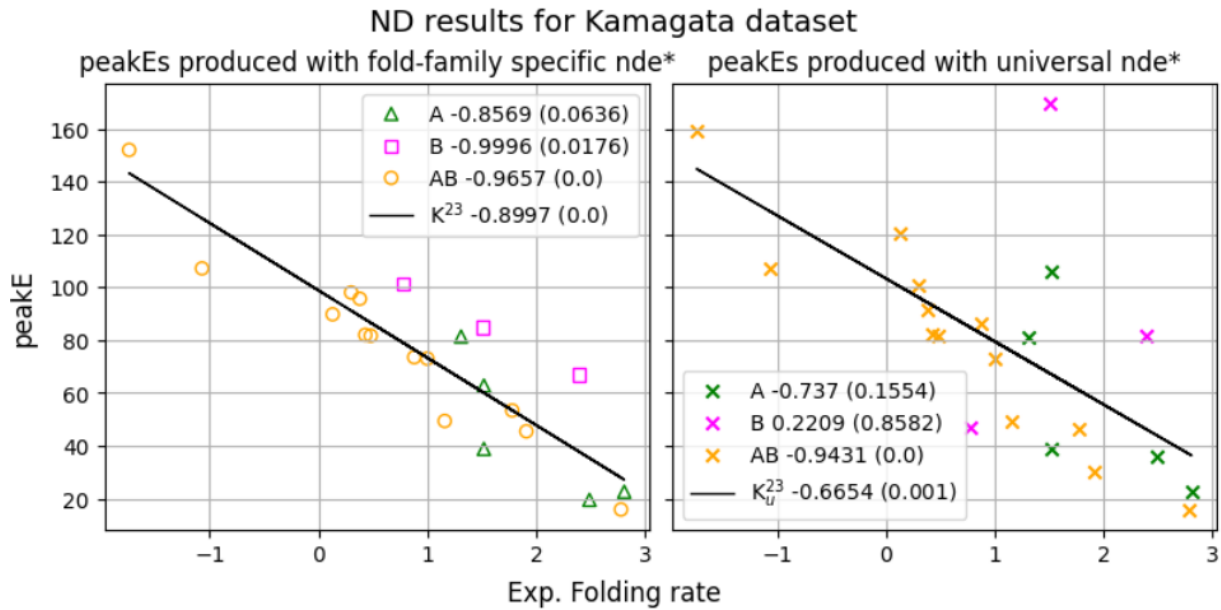

**Fig. 5** Left: ND results with random seed 23 and *nde** found by fold-family for the 21 multi-state folders in the Kamagata dataset. The combined Pearson's correlation coefficient is -0.8997 with p-value ≈ 0.0. Fold-family specific correlations are all below -0.856, with p-value < 0.064. Right: ND results with random seed 23 for the same dataset but *nde** found universally (for all chains in the dataset simultaneously). The combined Pearson's correlation coefficient is -0.6654 with p-value ≈ 0.001. Fold-family specific correlations are all below 0.221, with p-value < 0.859.

**Table 3.** The best *nde* policies by fold-family A, B and AB, and universally U, supporting the ND results in Fig. 5.

| Edge-state | H-H | H-HS | H-HT | H-HST | S-S | S-HS | S-ST | S-HST | T-T | T-HT | T-ST | T-HST | HS-HS | HS-HT | HS-ST | HS-HST | HT-HT | HT-ST | HT-HST | ST-ST | ST-HST | HST-HST |
|---|---|---|---|---|---|---|---|---|---|---|---|---|---|---|---|---|---|---|---|---|---|---|
| A | su | - | u | su | - | - | - | - | u | sl | - | - | - | - | - | - | u | - | u | - | - | u |
| B | - | - | - | - | u | - | s | - | su | - | su | - | - | - | - | - | - | - | - | sl | - | - |
| AB | u | u | u | u | u | su | u | u | u | u | u | u | u | u | u | u | u | u | u | s | u | u |
| U | su | u | u | u | s | u | u | u | s | s | u | s | u | sl | u | u | u | u | u | s | u | u |

The results shown in Figs. 4 and 5 were obtained with random seeds 13 and 23 for the Uzunoglu and Kamagata datasets, respectively. This element of serendipity indicates the importance of both the initial search point (parent genotype) and the prevailing condition provided by the random seed, towards locating an appropriate edge recovery policy (*nde**) efficiently. The two factors, suitable policies and random seed, help create a favourable environment for modelling protein folding with ND. They could be viewed as analogous to near physiological conditions necessary for proteins to collapse to their native functional folds.

However, more important than the computational model or its optimal policies, is what the sequence of restored edges could reveal about protein folding pathways. To facilitate this goal, trajectory data are available[‡] for further analysis and model improvement.

---

[‡] github.com/pfwscn